# Kalman Filter and Wavelet Cross-Correlation for VHF Broadband Interferometer Lightning Mapping

Ammar Alammari [1,2], Ammar A. Alkahtani [1,*], Mohd Riduan Ahmad [2], Fuad Noman [1], Mona Riza Mohd Esa [3], Muhammad Haziq Mohammad Sabri [1], Sulaiman Ali Mohammad [3], Ahmed Salih Al-Khaleefa [2,4], Zen Kawasaki [5] and Vassilios Agelidis [1,6]

[1] Institute of Sustainable Energy (ISE), Universiti Tenaga Nasional (UNITEN), Selangor 43000, Malaysia; ammarengineer2014@gmail.com (A.A.); fuad.noman@uniten.edu.my (F.N.); haziqsahaja@gmail.com (M.H.M.S.); vasagel@elektro.dtu.dk (V.A.)

[2] Atmospheric and Lightning Research Laboratory, Center for Telecommunication Research and Innovation (CeTRI), Fakulti Kejuruteraan Elektronik dan Kejuruteraan Komputer (FKEKK), Universiti Teknikal Malaysia Melaka (UTeM), Melaka 76100, Malaysia; riduan@utem.edu.my (M.R.A.); ahmed.salih89@siswa.ukm.edu.my (A.S.A.-K.)

[3] IVAT, Sekolah Kejuruteraan Elektrik, Fakulti Kejuruteraan, Universiti Teknologi Malaysia (UTM), Johor 8300, Malaysia; mona@fke.utm.my (M.R.M.E.); sulaimanalimohammad@gmail.com (S.A.M.)

[4] Network and Communication Technology (NCT) Lab, Fakulti Teknologi & Sains Maklumat (FTSM), Universiti Kebangsaan Malaysia (UKM), Bangi 43600 UKM, Selangor, Malaysia;

[5] Graduate School of Engineering, Osaka University, 1-1 Yamadaoka, Suita, Osaka 565-0871, Japan; thunderstorm1949@gmail.com

[6] Department of Electrical Engineering, Technical University of Denmark, Anjer Engeluds Vej 1 Bygning 101A, 2800 Kgs. Lyngby, Denmark

* Correspondence: ammar@uniten.edu.my

**Abstract:** Lightning mapping systems based on perpendicular crossed baseline interferometer (ITF) technology have been developed rapidly in recent years. Several processing methods have been proposed to estimate the temporal location and spatial map of lightning strikes. In this paper, a single very high frequency (VHF) ITF is used to simulate and augment the lightning maps. We perform a comparative study of using different processing techniques and procedures to enhance the localization and mapping of lightning VHF radiation. The benchmark environment involves the use of different noise reduction and cross-correlation methods. Moreover, interpolation techniques are introduced to smoothen the correlation peaks for more accurate lightning localization. A positive narrow bipolar event (NBE) lightning discharge is analyzed and the mapping procedure is confirmed using both simulated and measured lightning signals. The results indicate that a good estimation of lightning radiation sources is achieved when using wavelet denoising and cross-correlations in wavelet-domain (CCWD) with a minimal error of 3.46°. The investigations carried out in this study confirm that the ITF mapping system could effectively map the lightning VHF radiation source.   Paper DOI: https://doi.org/10.3390/app10124238

## 1. Introduction

Lightning is a natural phenomenon that occurs in the atmosphere. When electrical discharges are generated, electromagnetic radiations over different ranges of frequencies are produced, usually extending from the Ultra-Low Frequency (ULF) through the Ultra-High Frequency (UHF) [1]. Electromagnetic fields are emitted either from cloud-to-ground (CG) strikes, intra-cloud (IC) lightning, and some in-cloud processes [2–6]. The physics behind lightning initiation suggested that lightning is produced when a considerable amount of warm air is up-drafted to a height where an extremely low temperature can cause particles of frozen ice to produce electric charge disconnections [7].

Lightning strikes can cause harm to human beings and extend effects to systems operating in the surrounding environment. These effects can be minimized if reliable lightning mapping and localization systems are deployed to enable accurate prediction and detection of lightning strikes [4], [8–11]. Lightning mapping provides sufficient information on detected CG flashes at a particular range of frequencies. The very high frequency (VHF) band has been important for lightning mapping and for studying the lightning process over the past four decades [12].

The VHF lightning mapping is carried out in two different ways: time of arrival (TOA) and direction of arrival (DOA) or what is generally known as interferometry (ITF). The VHF lightning mapping has been traditionally performed using the TOA approach and narrow-band ITF systems to locate the radiation sources [13–15]. The TOA uses the time difference of arrival (TDOA) technique which requires at least four to five antennae [6]. On the other hand, the narrowband ITF system uses closely spaced multiple baseline antennae (in order of n wavelength, $n\lambda$) to resolve the radiation measurement ambiguities and obtain the desired angular resolution [9]. The use of broadband ITF for lightning localization was first introduced by [2], and has been developed rapidly during the past few decades [16–20]. Unlike the narrowband ITF, the broadband ITF uses all frequency components which are then used to produce a better resolution of lightning discharges. In the latest development of ITF systems, an array of closely spaced antennae (three perpendicular antennae) are used to capture and map the lightning radiation into the two-dimensional mapping of sources [9]. Despite the significant improvements made to map lightning using ITF systems, the mapping accuracy remains a challenge.

In this paper, we exploit the effects of different parameters of well-established methods to estimate the lightning maps accurately. We propose a modified cross-correlation method to solve problems in the DOA calculation and to improve the lightning location accuracy. We examine various noise filtering to reduce the low and high noise frequency contents. Noisy lightning signals impact the cross-correlation performance and lead to incorrect estimation of phase difference. Different generalized cross-correlation techniques are used to estimate the TDOA in a segment-wise manner. These techniques include conventional time and frequency domain cross-correlation. However, these generalized cross-correlation methods are more sensitive to the embedded noise in the signal, which results in wrong TDOA estimations. Therefore, we extend the proposed framework to include the wavelet-based cross-correlation, which reduces the noise and improves the system location accuracy. We then compare the proposed technique to the conventional cross-correlation techniques. We also investigate the effect of interpolating the cross-correlation function in determining the time difference. To assess the performance of the proposed lightning mapping approach, we introduce a simulation technique to generate a broadband ITF lightning. In the simulation part, we examine the effectiveness of the proposed methods in a noisy environment, where two Gaussian noises with different SNRs (in dB) are added to the simulated lightning signals.

This paper is organized as follows. In Section 2, we discuss the background of the broadband ITF, the preprocessing of the original data, the related steps of lightning mapping estimation, and the simulation procedure of lightning data. Section 3 presents the obtained lightning mapping from the measured VHF radiation sources and also a comparison with the simulation result. This paper is concluded in Section 4.

## 2. Materials and Methods

This section introduces the broadband ITF structure and the related methods used for two-dimensional lightning mapping. Then, we present the proposed procedure for lightning data simulations.

### 2.1. Broadband Interferometer (ITF)

Figure 1 shows the proposed lightning mapping general procedures consisting of three stages: (1) Preprocessing to assess the VHF signal quality, by removing noise components and performing amplitude normalization. (2) A piecewise procedure is used to segment the lightning signals into overlapped sliding windows to perform the cross-correlations and calculate the phase difference from overlapping VHF signals time difference at each pair of antennae. Different cross-correlation techniques in time-domain (TD), frequency-domain (FD) and wavelet-domain (WD) were applied. (3) The lightning maps are derived by calculating the azimuth and elevation angles from the ITF geometry.

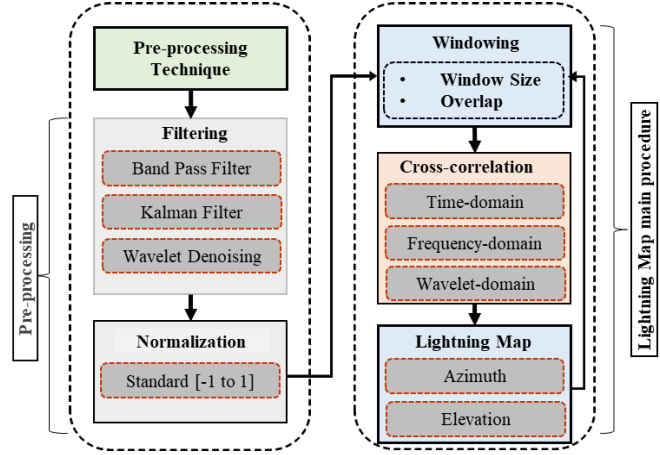

**Figure 1.** The procedure of lightning mapping.

2.1.1. Preprocessing

Prior to capturing the lightning electric field (E-field) signal at the VHF-ITF system, the lightning electric field signal travels from its origin in the clouds and usually propagates through open space. The signal is affected by reflection, refraction, absorption, and other interference means. To remove the unwanted noise from the recorded lightning, signals are usually filtered using bandpass filters with different cut-off frequencies [9,21,22]. In this paper, we investigate three types of noise filtering approaches including bandpass filter (BPF), Kalman filter (KF) [23,24], and wavelet transform (WT) denoising [25]. The key difference between these filters is that the BPF is the simplest and fastest approach of filtering in-band noises; however, it implicitly assumes that the noise occupies a specific frequency band (e.g., 20 to 100 MHz). Hence, the BPF fails if the noise occupies different frequency bands. The WT denoising solves this issue by decomposing the signal into different frequency sub-bands and apply thresholding rules to eliminate the unwanted frequency contents from different sub-bands. However, WT requires prior information of the specific bands when the noise exists. The Kalman filter can explicitly model the signal as a stochastic process given prior knowledge of the noise statistical parameters which make Kalman filter superior in cases where the noise frequency band is unknown. The main limitation of the Kalman filter is that it assumes the signal has only Gaussian noise.

2.1.2. Time Difference of Arrival (TDOA)

To calculate the TDOA and determine the direction of the lightning VHF radiation source, the cross-correlation has been traditionally used for estimating the similarities between two crossed baselines (consisting of three antennae). This process is achieved by using sliding overlapped windows. Besides, the generalized cross-correlation technique is one of the most classical time delay estimation methods.

Figure 2 illustrates the basic geometry of orthogonal baselines for two-dimensional lightning mapping. From Figure 2, the given two baselines (BC and BD) separated by distance *d* (15 m in this study), the TDOAs at each baseline can be calculated as,

$$\tau_{d1} = \frac{\Delta\phi_1}{2\pi f} \qquad (1)$$

$$\tau_{d2} = \frac{\Delta\phi_2}{2\pi f} \quad (2)$$

where $\tau_{d1}$ and $\tau_{d2}$ are the time differences of the baselines BC and BD, respectively. $\Delta\phi = 2\pi f \tau_d$ is the phase difference of a distant signal (with frequency $f$ in Hz) arriving at two antennae. $\Delta\phi$ can be found implicitly by performing the cross-correlation procedure.

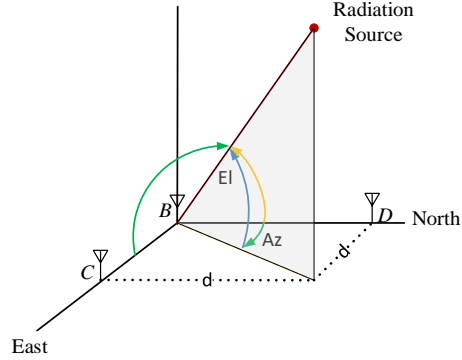

**Figure 2.** Basic interferometer geometry [9,19].

2.1.3. Lightning Mapping

For sources located in two spatial dimensions, two angles are needed to specify the direction and height of the radiation source, namely the azimuth ($Az$) and elevation ($El$) angles. From the spherical coordinate in Figure 2, the $Az$ and $El$ angles can be found by,

$$Az = \arctan\left(\frac{\tau_{d1}}{\tau_{d2}}\right) \quad (3)$$

$$El = \arccos\left(\frac{c}{d}\sqrt{\tau_{d1}^2 + \tau_{d2}^2}\right) \quad (4)$$

where $c$ is the speed of light, $\tau_{d1}$ and $\tau_{d2}$ are the TDOAs of the two orthogonal baselines. The TDOA ($\tau_d$) between every pair of antennae cannot exceed the transit time $\tau_{transit} = c/d$. This constrains the instance between the brackets of Equation (4) to have values in the range of ±1. However, any values exceed the ±1 range are considered as wrong estimations due to existing noise, and the corresponding $Az$ and $El$ angles can be neglected from the cross-correlation calculations.

In this paper, we compare the performance of three filtering approaches, a Butterworth BPF filter of order 4 and cut-off frequencies of (20–100 MHz) (fdesign MATLAB function), a Kalman filter (implemented in MATLAB [26] as in [24]), and a wavelet denoising technique (wdenoise MATLAB function). The wavelet filtering methods are abbreviated in this study for simplicity, e.g., WT-Coif5-SURE is the wavelet denoising using the Coiflet wavelet function of order 5 with Stein's unbiased risk estimate (SURE) thresholding rule. Besides the influence of different wavelet denoising functions, we also report the results of using two different thresholding rules, SURE and Universal Threshold. Wavelet denoising methods alter the estimated wavelet coefficients before the reversed transformation (reconstruction), which is achieved with some thresholding rules. The Kalman filter is a powerful tool that uses the state space formulations with a set of mathematical equations to recursively estimate future observations minimizing the mean square error of these estimates. Hence, the output of the Kalman filter is considered as a clean copy of the noisy signal where the filter successfully eliminates the Gaussian noises. Further details of the implemented Kalman filter in this paper can be found in [24].

Three cross-correlation methods are used; namely, cross-correlation in the time domain (CCTD), cross-correlation in the frequency domain (CCFD), and cross-correlation in the wavelet domain (CCWD). We have implemented the three cross-correlation methods in MATLAB, where a recursive sliding dot-product was used for CCTD, recursive conjugate dot-product of fft MATLAB function for CCFD, and modwtxcorr MATLAB function for CCWD. The cross-correlation is used to measure the similarity of lightning signals received at multiple antennae. In this context, a measure of similarity is to locate the peak of cross-correlation (the highest value of correlation coefficients over time lags) which then used to calculate the time delay of arrival between the correlated signals.

*2.2. Hardware*

Figure 3 shows the configuration diagram of the proposed ITF system which consists of three sets of antennae, with two perpendicular baselines of 15 m in length (3λ), band-pass filters, low noise amplifiers (LNA), coaxial cables, and an oscilloscope. Two types of parallel-plate antennae of size A4 and A3 are used to capture the VHF E-field and Fast E-field (10 Hz up to 3 MHz) of lightning flashes, respectively. For the VHF E-field antenna, the gap between the two plates is about $h$ = 1 cm and 3 cm for Fast E-field to avoid the fringing effects for the VHF sensor [26,27]. The Fast E-field antenna is used to determine the type of lightning flash, and the VHF E-field antenna is used to capture the lightning radiation emissions. Figure 3 illustrates the implemented antennae B, C, and D were arranged in perpendicular baselines with two sides BC and BD to measure the phase difference. The broadband ITF system was installed in Klebang Beach, Malacca Strait, Melaka, Malaysia (2.216304° N, 102.192288° E).

The received VHF signal at each antenna is sequentially passed through preamplifier and BPF before being digitized and stored in external memory. The LNA amplifies the low-power signals preserving the signal-to-noise-ratio of the received signal. The bandwidth of the BPF is limited to the range 40–80 MHz with a center frequency of 60 MHz [17]. The filtered analog signal is then passed to the oscilloscope (PicoScope 3000 series [29] with 100 MHz bandwidth) through three coaxial cables of 15 m length. The output of the antenna was digitized in the oscilloscope at a sampling frequency of 250 MS/s with a vertical resolution of 8 bits. Data records were event-triggered and stored in segments of 200 ms length. Then, the retrieved data from the oscilloscope are transferred to a personal computer with MATLAB software for lightning localization analysis. A summary of the ITF's attributes is given in Table 1.

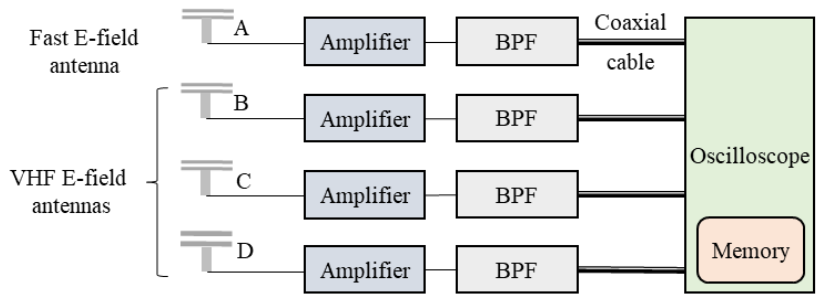

**Figure 3.** VHF broadband digital interferometer data acquisition system.

*2.3. Simulation*

In this sub-section, numerical simulations are conducted to assess the performance of the proposed mapping methods using simulated VHF-ITF signals. An actual measured signal (from antenna B is used as a reference to simulate the other perpendicular antennae (C and D). Given the hand-crafted azimuth and elevation incident angles, the signals of antennae C and D are copies of antenna B with a phase shift for each temporal sliding window. This simulation can provide an insight into the best methodological approach for estimating lightning mapping. Figure 4 illustrates the proposed steps to simulate the lightning mapping signals and to evaluate the performance of the estimation methods.

**Table 1.** The parameters for the interferometer and the lightning flash.

| Parameters | Description |
|---|---|
| Sample interval | 4 ns |
| Sampling Rate of VHF | 250 MS/s |
| No of samples | 52,000,002 |
| Number of Antennae | 3 |
| Signal band-limited | 40 MHz-80 MHz |
| Centred Frequency | 60 MHz |
| Distance between antennae | 15 m baseline |
| Picoscope | Picoscope 3000 (3405D) |

2.3.1. Generation of Azimuth and Elevation Angles

This part presents some basic but powerful data augmentation techniques. Data augmentation is a strategy that enables practitioners to significantly increase the diversity of data available for analysis, without actually collecting new data.

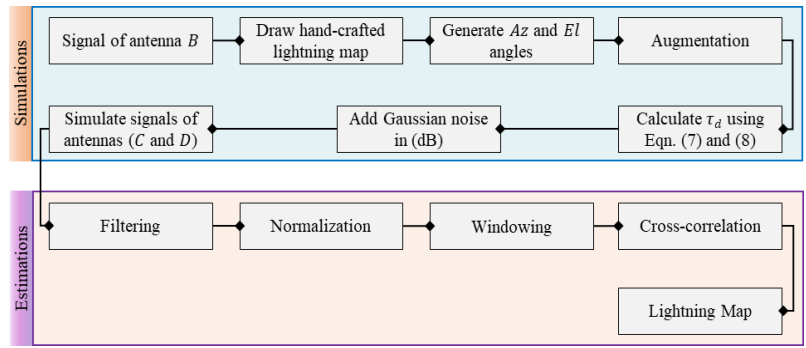

**Figure 4.** The procedure of simulating lightning signals.

Various data augmentation techniques were implemented in this study, such as adding Gaussian noise, scaling, and flipping and their combinations. Figure 5 shows some examples of these lightning data augmentation techniques. The generated $Az$ and $El$ angles from the hand-crafted lightning signal are assumed to be real-positive or negative values. The Gaussian noise $\varepsilon$ is an independent and identically distributed random variable, $N \sim (0,1)$, with zero-mean and unit-variance, which is added to the $El$ angles; the output $\hat{E}l_i = El_i + \varepsilon_i, \forall \tau_{di} < \tau_{transit}$ is a shifted version of the real $El$ values constrained to the transit time between two antennae $\tau_{transit} = \frac{d}{c} = 5 \times 10^{-8}\ s$.

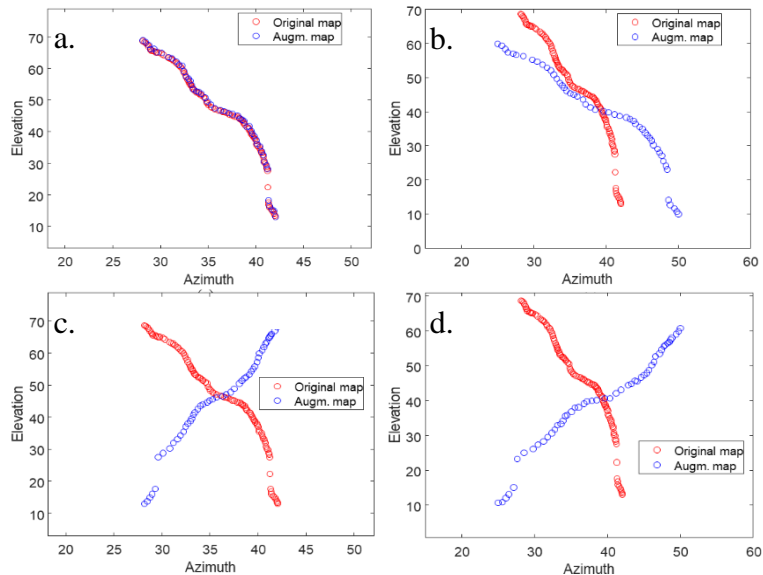

**Figure 5.** Example of angles augmentation by (**a**) adding Gaussian noise, (**b**) scaling, (**c**) flipping, (**d**) combination of (a, b, and c).

For scaling augmentation, since the simulated azimuth ($Az$) and elevation ($El$) angles fall in a relatively narrow range, the outward scaling is used as it expands the azimuth and elevation range, forcing the proposed methods to make assumptions about what lies beyond the simulated values boundary. The lightning representations can be flipped horizontally and vertically. However, in practice, lightning mapping is not preferred to be flipped vertically. A horizontal flip is equivalent to rotating the lightning curve by 180 degrees.

2.3.2. Derivation of Time Delay

For calculating the time delay $\tau_{d1}$ and $\tau_{d2}$, a reverse mathematical derivation of the geometrical equations in Subsection 2.1.2 is followed. $\tau_{d1}$ can be derived from the spherical Equation (3) and as follows,

$$\tau_{d1} = \tan(Az)\, \tau_{d2} \tag{5}$$

From Equations (4) and (5), the TDOA $\tau_{d2}$ can be found by,

$$\cos(El) = \frac{c}{d}\sqrt{\tau_{d1}^2 + \tau_{d2}^2}$$

$$\frac{d}{c}\cos(El) = \sqrt{\tau_{d1}^2 + \tau_{d2}^2}$$

$$\frac{d^2}{c^2}\cos^2(El) = \tau_{d1}^2 + \tau_{d2}^2 \tag{6}$$

Substituting Equation (5) into Equation (6) we get

$$\frac{d^2}{c^2}\cos^2(El) = (\tan(Az)\tau_{d2})^2 + \tau_{d2}^2$$

$$\frac{d^2}{c^2}\cos^2(El) = \tan^2(Az)\, \tau_{d2}^2 + \tau_{d2}^2$$

$$\frac{d^2}{c^2}\cos^2(El) = \tau_{d2}^2((1 + \tan^2(Az))$$

$$\tau_{d2}^2 = \frac{d^2 \cos^2(El)}{c^2((1 + \tan^2(Az))}$$

$$\tau_{d2} = \sqrt{\frac{d^2 \cos^2(El)}{c^2((1 + \tan^2(Az))}} \tag{7}$$

Consequently, substituting Equation (7) into Equation(5), we can obtain the time difference between the two antennae:

$$\tau_{d1} = \tan(Az)\sqrt{\frac{d^2\cos^2(El)}{c^2(1 + \tan^2(Az)}}$$

$$\tau_{d1} = \frac{d \tan(Az) \cos(El)}{c\sqrt{(1 + \tan^2(Az))}} \tag{8}$$

2.3.3. Simulated ITF Signals

Given a captured lightning signal of antenna $B$ (located at Cartesian coordinate 0 of Figure 2), Figure 6 illustrates the proposed simulation steps, including windowing, calculating the amount of temporal shift for each particular window, and using Equations (7) and (8) to produce corresponding simulated step-wise signals for antennae *C* and *D,* respectively. The simulated sliding windows are then concatenated to form a full track of VHF-ITF lightning signals.

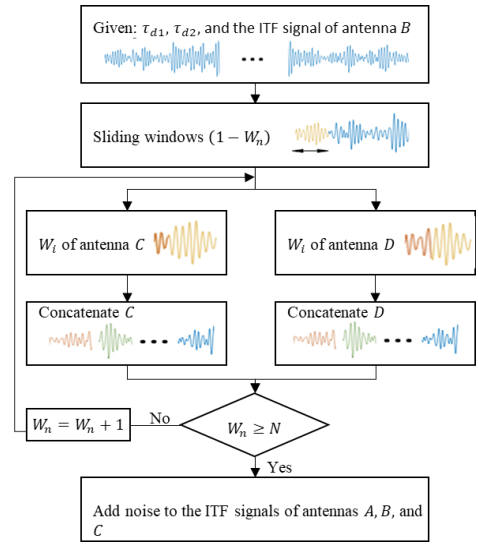

**Figure 6.** The general procedure of Simulating ITF Signal. Where $W_n$ is the window length (i.e., 256 samples), $W_i, i = 1,\ldots,N$ is window counter, $N$ is the total number of sliding windows of the processed ITF signal, $W_n + 1$ is the temporal shift for subsequent windows (1 sample).

## 3. Results and discussion

In this paper, we consider the use of the broadband ITF technique for the analysis of VHF lightning mapping. A comparison of various processing methods of the VHF lightning signals is performed. The simulation was conducted to investigate the impact of noise and data resolution on lightning mapping accuracy. Different filtering and cross-correlation techniques were implemented introducing new processing methods such as Kalman filter and wavelet-based cross-correlation. The best performing set of methods are validated using collected real lightning signals. Figure 7 shows an example of cross-correlation steps to determine the TDOA $\tau_d$ using two antennae sliding windows.

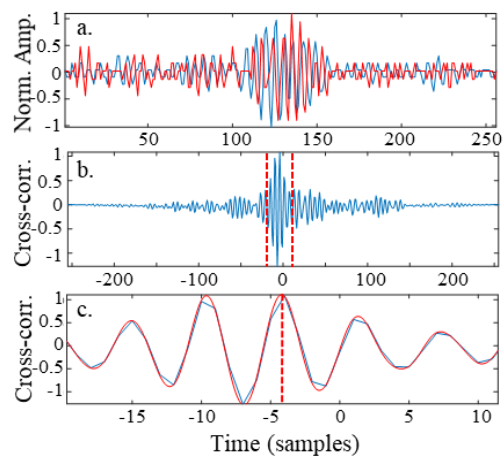

**Figure 7.** Example of the processing steps used to determine the TDOA ($\tau_d$) at antennae B and C. (**a**) Normalized waveforms from the two antennae B and C for a 256 (1.042 μs) sample window; (**b**) the cross-correlation of antennae B and C signals with two vertical red dash-lines indicate the zoomed plot area in c; (**c**) illustrating the interpolation fitting (red color) of the expanded view cross-correlation (blue color) with vertical red dash-line showing the peak of correlation.

*3.1. Simulation Results*

The performance of the lightning mapping estimation was measured based on the number of sliding windows over the time course of the analyzed ITF lightning signals, where each pair of $Az - El$ angles are correctly estimated relative to the ground-truth simulated lightning map. The Euclidean distance between the estimated and the actual-simulated lightning map was used to measure the estimation error,

$$dist = \frac{1}{W_n} \sum_{i=1}^{W_n} \sqrt{(Az_i - \hat{A}z_i)^2 + (El_i - \hat{E}l_i)^2} \quad (9)$$

where $W_n$ is the total number of sliding windows in the processed lightning data. To summarize the performance of the simulated dataset overall, we compute the average of Euclidean distances.

Table 2 summarizes the performance comparisons of the suggested lightning mapping estimation methods. In Table 2, The Euclidean distances were calculated for each simulated ITF datum and then averaged over the entire available augmented lightning dataset to show the overall benchmarking performance of the suggested approaches. We compare three main filtering approaches listed in the first column of Table 2. It can be seen that, when using the CCTD method, the KF and BPF show the lowest Euclidean distances of 3.55° and 3.63° respectively. The linear-interpolation factor used to achieve these results was 8. The WT denoising also shows relatively low distances of 3.94° when using the Fejer-Korovkin function with cubic-interpolation of factor 8.

For the CCFD method, the best performance was achieved using the Kalman filter without interpolation with a distance of 3.59°, while the BPF and wavelet denoising performed comparably to that of the CCTD method showing minimum distances of 3.76° and 3.94°, respectively. On the other hand, using CCWD, the WT denoising shows the best performance with a distance of 3.46° using the Symlet function without any interpolations. BPF and KF also provide comparable results to that of the WT with 3.76° and 3.47°.

Generally, from Table 2, the filtering methods do not show significant enhancements in the performance when increasing the cross-correlation interpolation factor due to the relatively small window size adapted from Stock et al. [9]. The introduced wavelet filtering shows better performance than that of the BPF. However, the variation of wavelet functions and the thresholding methods show almost similar performances. On the other hand, the KF outperforms both the BPF and the WT in general. This is because KF can recursively estimate the noise regardless of the frequency contents.

In comparing the cross-correlation techniques, the CCWD method outperforms the other methods in all cases. This is due to the inherent capability of wavelet transforms to enhance the signals further and reduce the noise effects, providing a better time-domain resolution and reliable estimation of lighting maps. Moreover, comparing several wavelet functions in WT denoising helps to identify the best combination of WT parameters that fits the analyzed signals. A noticeable enhancement of cross-correlation methods was found introducing cross-correlation in the wavelet domain for estimating the lightning maps.

**Table 2.** Comparison (Euclidean distance) of lightning mapping estimation consisting of different filtering and cross-correlation approaches.

| Correlation method | CCTD | | | | | | | | CCFD | | | | | | | | CCWD | | | | | | | |
|---|---|---|---|---|---|---|---|---|---|---|---|---|---|---|---|---|---|---|---|---|---|---|---|---|
| Interp. Method | Linear | | | | Cubic | | | | Linear | | | | Cubic | | | | Linear | | | | Cubic | | | |
| Interp. Factor | 1 | 2 | 4 | 8 | 1 | 2 | 4 | 8 | 1 | 2 | 4 | 8 | 1 | 2 | 4 | 8 | 1 | 2 | 4 | 8 | 1 | 2 | 4 | 8 |
| WT-Coif5-SURE | 4.08 | 3.99 | 4.04 | 4.10 | 4.08 | 4.16 | 4.10 | 4.08 | 4.08 | 4.35 | 4.53 | 4.60 | 4.08 | 4.16 | 4.10 | 4.08 | 3.65 | 3.90 | 3.86 | 4.76 | 3.65 | 4.80 | 4.78 | 4.75 |
| WT-Coif-UNIVERSAL | 8.54 | 8.23 | 8.24 | 8.29 | 8.54 | 8.40 | 8.37 | 8.35 | 8.54 | 8.65 | 8.85 | 8.95 | 8.54 | 8.40 | 8.37 | 8.35 | 11.22 | 11.51 | 11.84 | 12.65 | 11.22 | 12.67 | 12.62 | 12.59 |
| WT-db10-SURE | 4.19 | 4.10 | 4.08 | 4.16 | 4.19 | 4.10 | 4.07 | 4.03 | 4.19 | 4.41 | 4.47 | 4.55 | 4.19 | 4.10 | 4.07 | 4.03 | 3.85 | 3.89 | 3.79 | 4.62 | 3.85 | 4.57 | 4.57 | 4.54 |
| WT-db10-UNIVERSAL | 9.10 | 8.98 | 9.05 | 9.12 | 9.10 | 9.03 | 9.00 | 8.97 | 9.10 | 9.15 | 9.27 | 9.32 | 9.10 | 9.03 | 9.00 | 8.97 | 11.39 | 11.66 | 11.59 | 12.23 | 11.39 | 12.21 | 12.23 | 12.19 |
| WT-FK14-SURE | 4.06 | 3.96 | 3.95 | 4.02 | 4.06 | 3.99 | 3.96 | 3.94 | 4.06 | 4.30 | 4.37 | 4.44 | 4.06 | 3.99 | 3.96 | 3.94 | 3.98 | 4.10 | 3.92 | 4.80 | 3.98 | 4.80 | 4.68 | 4.65 |
| WT-FK14-UNIVERSAL | 9.30 | 9.24 | 9.35 | 9.44 | 9.30 | 9.26 | 9.24 | 9.22 | 9.30 | 9.34 | 9.47 | 9.53 | 9.30 | 9.26 | 9.24 | 9.22 | 12.40 | 12.47 | 12.30 | 12.96 | 12.40 | 12.85 | 12.95 | 12.92 |
| WT-Sym4-SURE | 4.06 | 3.97 | 4.00 | 4.07 | 4.06 | 4.11 | 4.07 | 4.03 | 4.06 | 4.36 | 4.50 | 4.62 | 4.06 | 4.11 | 4.07 | 4.03 | 3.46 | 3.74 | 3.71 | 4.59 | 3.46 | 4.62 | 4.57 | 4.51 |
| WT-Sym4-UNIVERSAL | 10.46 | 10.18 | 10.17 | 10.23 | 10.46 | 10.31 | 10.31 | 10.27 | 10.46 | 10.60 | 10.72 | 10.84 | 10.46 | 10.31 | 10.31 | 10.27 | 11.84 | 12.22 | 12.37 | 13.16 | 11.84 | 13.20 | 13.14 | 13.10 |
| BPF | 3.76 | 3.63 | 3.77 | 3.85 | 3.76 | 3.83 | 3.80 | 3.76 | 3.76 | 3.97 | 4.17 | 4.27 | 3.76 | 3.83 | 3.80 | 3.76 | 3.76 | 3.86 | 3.76 | 4.62 | 3.76 | 4.62 | 4.59 | 4.54 |
| KF | 3.59 | 3.55 | 3.65 | 3.84 | 3.59 | 3.71 | 3.80 | 3.61 | 3.59 | 3.89 | 4.07 | 4.25 | 3.59 | 3.71 | 3.80 | 3.61 | 3.47 | 3.78 | 3.68 | 4.55 | 3.47 | 4.53 | 4.51 | 4.48 |

*3.2. Real Data Results*

The main challenge of source localization is to identify the radiation emitted by the source out of noisy signals received by the measuring ITF system. Figure 8 shows an example of a positive narrow bipolar event (NBE). Figure 9 shows measured VHF signals related to a lightning flash triggered by the discharge process of the NBE, as observed in Figure 8. This measurement represents a time expanded window of 20 µs (5000 samples). The observed signals of antenna B and D show significant similarity with possible phase shift or a time lag. However, the signal of antenna C shows slightly different noise characteristics.

In the simulation results shown before, it is noted that the wavelet-based CCWD method produced the best performance of lightning mapping using the Symlet function with linear interpolation of level 8. Therefore, the same setup is implemented for real lightning data mapping.

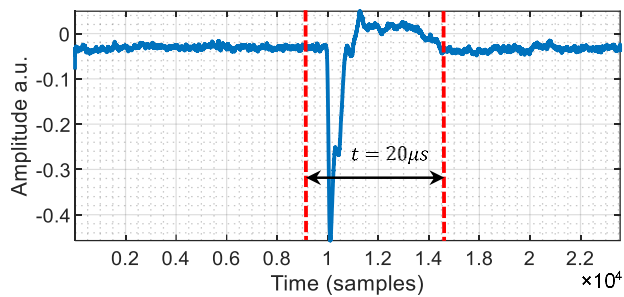

**Figure 8.** Fast Electric field pulse relevant to a positive NBE.

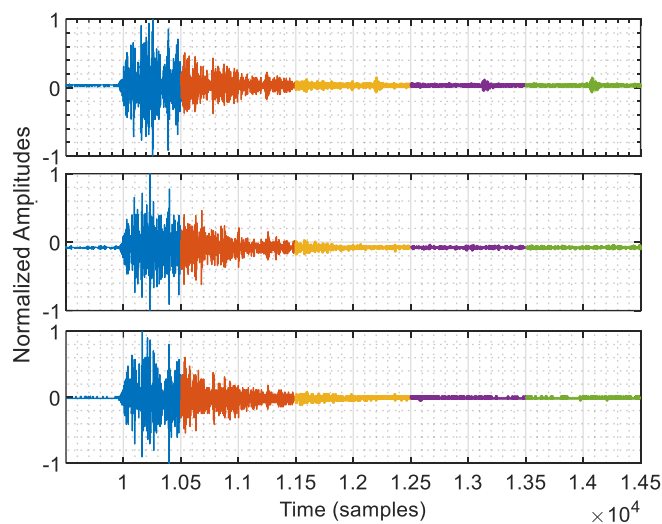

**Figure 9.** The VHF radiation signals associated with positive NBE received by antennae B, C, and D.

Figure 10 shows the constructed map of the lightning source location for positive NBE. The starting position of radiation sources was in the higher altitude about 66° elevation. However, these detected sources are isolated from the well-defined lightning channel and considered outliers. The colour changes in Figure 10a with time from blue to green, corresponding to the lightning progression time in Figure 9. It can be seen that the channel from the bottom (23° elevation) to the 32° elevation was approximately straight with small changes in azimuth angles range, corresponding to the trajectory of the lightning flash. Up to the 32° elevation, the shape and the height of the channel showed certain consistency with the observation of positive NBE. In Figure 10b, the radiation elevation angles and fast electric field change observations (black waveform) are superimposed on the VHF waveform (gray), showing the downward propagation of the VHF source. The negative electric field waveform is indicative of an upward-directed current, consistent with a positive NBE. Overall, all three filtering methods implemented in this paper show comparable accuracy for this lightning mapping with some errors that could result from the limited windowing approach conducted in this study, or due to the existing in-band hardware-interference noises when collecting the data. To avoid such noise effects, it is better to properly filter the analog signal from the antennae before the signal is digitized, such that extreme noises contaminate no portion of the bandwidth which contributes to the cross-correlation.

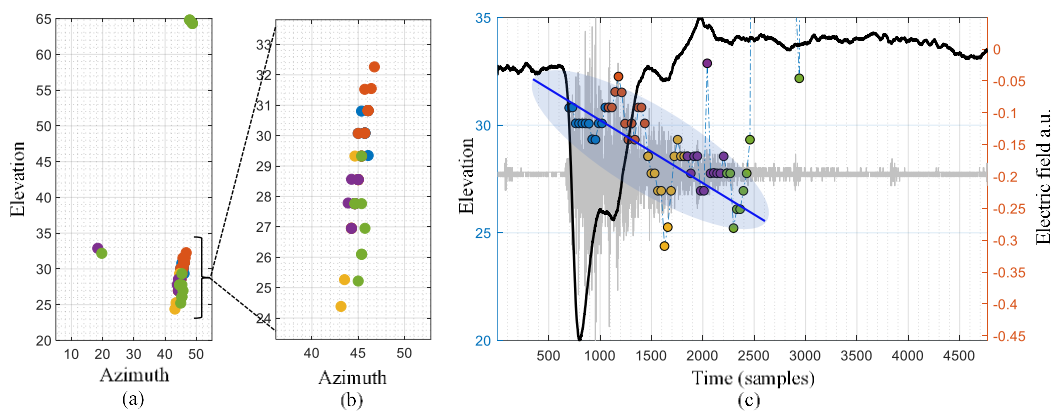

**Figure 10.** Interferometer data for NBE. (**a**) Lightning map plotted in elevation vs. azimuth. (**b**) Expanded view of the lightning map of a. (**c**) Showing the breakdown, each colored circle-marker denotes the elevation angles (altitude) vs. time.

## 4. Conclusion

Broadband interferometry for the two-dimensional mapping of lightning progression was used to examine the effects of different factors on the accuracy of the lightning maps. In this paper, various lightning signal proceeding approaches are compared, including different filtration methods and cross-correlation techniques, introducing interpolation to effectively estimate the fractional time of arrival delay realizing the location of VHF lightning radiation sources in two dimensions. Simulations were conducted to provide enough analysis when comparing the performance of implemented approaches for estimating lightning maps. Aggregated results show that the CCWD outperforms the other correlation methods in estimating the TDOAs regardless of the interpolations. This is because the wavelet transform not only can reduce the noise but also weakens the effects of the peak resolution of cross-correlation. The filtering methods also show comparable performance with wavelet denoising, which shows slightly better performance when using Symlet function. Several improvements of the current study could be achieved by involving other parameters, such as tuning the sliding window size, the windowing function, the amount of overlap.

**Author Contributions:** Conceptualization, A.A. and M.R.A.; methodology, A.A.; software, A.A. and F.N.; validation, A.A.A., M.R.A., Z.K. and M.R.M.E.; formal analysis, A.A.; investigation, M.H.M.S.; resources, M.H.M.S. and S.A.M.; data curation, M.H.M.S.; writing—original draft preparation, A.A.; writing—review and editing, A.S.A.-K.; visualization, V.A.; supervision, M.R.A. and A.A.A.; project administration, M.R.A.; funding acquisition, A.A.A. and M.R.A. All authors have read and agreed to the published version of the manuscript.

**Funding:** This work was supported by Universiti Tenaga Nasional (UNITEN) under BOLD2025 fund, in part by the Ministry of Higher Education, Malaysia, under Fundamental Research Grant Scheme (FRGS) (FRGS/2018/FKEKK-CETRI/F00361), and in part by Universiti Teknologi Malaysia (UTM) under Q04G19 and R4F966 research grants (FRGS)..

**Conflicts of Interest:** The authors declare no conflict of interest. The funders had no role in the design of the study; in the collection, analyses, or interpretation of data; in the writing of the manuscript, or in the decision to publish the results.